\documentclass[aps,prl,preprint,groupedaddress]{revtex4}
\usepackage[german,american]{babel}

\usepackage[final]{graphicx}% Include figure files
\usepackage{amssymb}

\makeatother

\newcommand{\dwp}{DWP}
\newcommand{\dwpnd}{DWP$^{\mathrm{ND}}$}
\newcommand{\dwpod}{DWP$^{\mathrm{WD}}$}
\newcommand{\dwpli}{DWP$^{\mathrm{Li}}$}
\newcommand{\tlsnd}{TLS$^{\mathrm{ND}}$}
\newcommand{\tlsod}{TLS$^{\mathrm{WD}}$}
\newcommand{\tlsli}{TLS$^{\mathrm{Li}}$}

\newcommand{\lj}{LJ}
\newcommand{\bmlj}{BMLJ}
\newcommand{\ang}{\ensuremath{\text{\AA}}}

\begin{document}

%\preprint{APS/123-QED}

\title{Microscopic description of the low-temperature anomalies in silica and lithium silicate via computer simulations}

\author{J. Reinisch}
 %\\altaffiliation[Also at ]{Physics Department, XYZ University.}%Lines break automatically or can be forced with \\
\author{A. Heuer}%
 %\\email{Second.Author@institution.edu}

\affiliation{
Westf\"{a}lische Wilhelms-Universit\"{a}t M\"{u}nster, Institut f\"{u}r Physikalische Chemie\\
Corrensstr. 30, 48149 M\"{u}nster, Germany }

\date{\today}% It is always \today, today,

\begin{abstract}
Information about the nature of the low-temperature anomalies and in
particular the properties of the tunneling systems in silica and
lithium silica glasses are revealed via computer simulations. The
potential energy landscape of these systems is systematically
explored for adjacent pairs of local minima which may act as
double-well potentials (DWP) at low temperatures. Three different
types of DWP are distinguished, related to perfectly coordinated
silica, intrinsic silica defects, and extrinsic defects. Their
properties like the spatial extension and the dipole moment are
characterized in detail. Furthermore, the absolute number of
tunneling systems, i.e. symmetric DWP, is estimated. The results are
compared with dielectric echo, specific heat and acoustic
experiments on Suprasil I and Suprasil W. A semi-quantitative
agreement for all relevant features is obtained.
\end{abstract}

\pacs{0.8.15}% PACS, the Physics and Astronomy
\maketitle

\section{\label{introduction}INTRODUCTION}
For more than thirty years it is known that glasses display
anomalous behavior for temperatures in the Kelvin regime.$^1$ For
example, the temperature dependence of the specific heat turns out
to be linear rather than cubic, as expected from the Debye model.
The standard tunneling model (STM)
 predicts many of these features.$^{2,3}$
The key idea is to assume that the system can perform local
transitions between adjacent configurations. Because of the low
temperatures the crossing of the barriers occurs via tunneling. This
feature can be characterized by a double-well potential (DWP) in
configuration space, i.e. two adjacent local minima on the potential
energy landscape (PEL). In general, the reaction coordinate
corresponds to a highly cooperative process in which a group of
neighboring atoms is involved. Among all DWP which are present in a
given glassy configuration only those DWP with asymmetries less
than, let`s say, 2 K will contribute to the low-temperature
anomalies. In contrast, typical energy scales for, e.g.,  SiO$_2$
are of the order of 1eV and thus many orders of magnitude higher.
Thus one would expect that only a minor fraction of all DWP are
relevant for the low-temperature anomalies. They are denoted
two-level systems (TLS).

There is no theory which predicts the properties of TLS from first
principles without invoking some model assumptions. There are,
however, interesting approaches within different physical frameworks
like a mean-field model $^4$ and random first order transition
theory.$^5$ In the second approach a detailed picture of the PEL in
terms of mosaics is derived. In their model the reaction coordinate
of TLS involves the dynamics of $O(10^2$) molecules which only move
a small fraction of a nearest-neighbor distance. Finally, in the STM
some reasonable assumptions about the properties of the potential
parameters of the TLS are made.

Computer simulations are well suited to elucidate the properties of
TLS. Two steps are involved. First, one has to generate a typical
glassy configuration and, second, to identify nearby adjacent minima
on the PEL and characterize the nature of the individual
transitions. Qualitatively, one would expect that the active atoms
for one transition are localized in some region in the glass. The
first simulations used non-systematic search methods.$^{6,7}$ In
subsequent work Heuer and Silbey have introduced a method how to
search TLS in a systematic way.$^{8}$ Applying this method to a
binary mixture Lennard-Jones (BMLJ) system  they were able to
predict the real-space realization of the TLS $^{8}$, the coupling
to strain, $^{9,10}$ the prediction of tunneling properties in terms
of material constants and, as a consequence, a universal description
of the low-temperature anomalies.$^{11,12}$ Important results about
the properties of DWP in Lennard-Jones systems can be also found in
Refs. $^{13-15}$. For all simulations one has to take into account
that the absolute number of TLS is so small that there is no chance
to identify a sufficiently large number via simulations. In
contrast, the number of DWP is very large. Therefore an important
ingredient of these simulations is the formulation of a statistical
method how to predict the properties of TLS from those of DWP. Using
the significantly improved computer facilities during the last
decade, the present authors have recently reanalyzed the BMLJ system
.$^{16,17}$ In particular, it could be shown that (1) the systematic
search algorithm indeed allows one to obtain an estimation of the
absolute number of TLS, (2) the properties of the TLS basically do
not depend on the energy of the configuration. Thus, the intrinsic
computer limitations like the necessity of a very fast cooling
procedure to generate the initial configuration or the choice of
relatively small systems do not hamper the unbiased determination of
TLS properties in the BMLJ system.

A prototype glass-forming system is vitreous silica (SiO$_2$).
Trachenko et al have characterized the TLS of silica in several
respects and have shown that the microscopic nature is related to
coupled rotations of SiO$_4$ tetrahedra.$^{18-21}$ A similar picture
has been suggested from experimental neutron data $^{22}$ and from
the analysis of localized vibrational modes. $^{23}$ The transition
events have been also analyzed via the activation-relaxation
technique.$^{24}$ The present authors have conducted a systematic
search for TLS in silica.$^{25}$ It has been possible to estimate
the experimentally observed number of TLS which has turned out to be
a factor of approx. 3 smaller than the experimental TLS density
reported for Suprasil W. As argued in Ref.$^{25}$ there are two
possible reasons for a slight underestimation by computer
simulations which are connected with the estimation of the tunneling
matrix element. Going beyond the WKB-approximation one might
additionally take into account that (i) the saddles are broader (in
terms of second derivatives) than the minima (see also
Refs.$^{14,17}$) and (ii) there is a Franck-Condon factor, taking
into account the relaxation of the phonon modes during the
transition within a DWP.$^{17}$ Both effects would increase the
estimation of the number of TLS and might at least partly account
for the remaining difference.

In general, one may distinguish intrinsic and extrinsic TLS.$^{26}$
The extrinsic TLS may be related to extrinsic defects like
OH-impurities. Actually, comparing Suprasil I (1250 ppm
OH-impurities) with Suprasil W (< 5 ppm OH-impurities) the
experimentally determined number of TLS is nearly twice as high.
This may be related to the additional contribution of extrinsic
TLS.$^{26}$ For the intrinsic TLS two different contributions may be
of relevance. First, they may be related to TLS where {\it
intrinsic} defects of the silica system are involved. They are
characterized by non-tetrahedral local coordination like
non-bridging oxygen atoms. Second, intrinsic TLS may also be present
in defect-free silica configurations which recently have been
analyzed in Ref.$^{25}$. Intuitively, one might expect that
defective configurations (intrinsic and extrinsic) are more
efficient to locally reorganize and thus to form TLS. Both, the
presence of intrinsic and extrinsic defects, will be relevant for a
closer rationalization of the low-temperature anomalies in systems
like Suprasil.

In previous work on a BMLJ system the properties of extrinsic
defects have been analyzed in detail.$^{27}$ For this purpose one
has included a test Lennard-Jones particle with variable radius and
analyzed the DWP, related to this particle. Some dramatic effects
have been observed when analyzing a test particle with a radius
which is e.g. 25\% smaller than the radius of the smaller component
of the BMLJ system (data compared with the properties of the
majority component): (i) the probability for the formation of a DWP
has increased by a factor of approx. 50, (ii) the number of
particles, involved in the DWP, has decreased by more than a factor
of 2, (iii) the average saddle height of the DWP has increased by a
factor of nearly 2, (iv) the deformation potential has decreased by
25\%.

In this work we explicitly compare the nature of the TLS of
defect-free silica with those of silica containing intrinsic
defects and/or extrinsic defects, modelled by a small
concentration of Li$_2$O. It will turn out that all types of TLS
have very different properties.  On a qualitative level, many
features for the extrinsic defects will turns out to be similar to
what has been observed for the test particle in the BMLJ system,
as sketched above. To underline the interpretation of the results,
we compare them with the respective observations for the BMLJ
system and with experimental data.

\section{\label{technical}TECHNICAL}
The molecular dynamics (MD) simulations have been conducted under
NVE-conditions. The velocity Verlet scheme has been chosen to
propagate the particles in time. The MD simulations were basically
used to generate large numbers of independent structure, which then
were analyzed for the occurrence of \dwp. Periodic boundary
conditions have been applied.

\subsection{Binary mixture Lennard-Jones}
As a \lj-model glass former we chose a binary mixture system with
80\% large A-particles and 20\% small B-particles (\bmlj).$^{28-31}$
It is supposed to represent NiP (80\% $^{62}$Ni; 20\% $^{31}$P)
 but with a 20\% higher particle density.$^{32}$
This system was first used by Kob and Anderson. The used potential
is of the type
\begin{equation}
V_{\alpha\beta}=4\cdot\epsilon_{\alpha\beta}[(\sigma_{\alpha\beta}/r)^{12}-(\sigma_{\alpha\beta}/r)^{6}]
  +(a +b\cdot r),
\end{equation}
with $\sigma_{AB}=0.8\sigma_{AA}$, $\sigma_{BB}=0.88\sigma_{AA}$,
$\epsilon_{AB}=1.5\epsilon_{AA}$, $\epsilon_{BB}=0.5\epsilon_{AA}$,
$m_B = 0.5 m_A$. A linear function $a + b \cdot r$ was added  to
ensure continuous energies and forces at the cutoff $r_c=1.8$.  The
units of length, mass and energy are $\sigma_{AA}$, $m_A$,
$\epsilon_{AA}$, the time step within these units was set to $0.01$.
For the case of NiP the energy unit corresponds to 934 K and
$\sigma_{AA}$ is 2.2\ {\AA}. The presented data are taken from our
smallest simulated system with 65 particles. The finite size effects
have already been analyzed in Ref.$^{16,17}$ and are small enough to
be neglected.

\subsection{Silica}
The pure silica system has been modeled by the BKS-potential.$^{33}$
We have analyzed system sizes of N=150 and N=600 particles. If not
mentioned otherwise the data in this work refer to $N=150$. The
system has the standard density of 2.3 $\mathrm{g/cm^3}$ and a
short-range cutoff of the BKS-potential of 8.5 \AA. The starting
configurations for the systematic search correspond to equilibrium
configurations at 3000 K, which subsequently were minimized. We have
obtained starting structures with and without intrinsic defects,
i.e. deviations from a perfect tetrahedral coordination. Due to the
fact, that defect related \dwp\ may be relevant, though the number
of defects is very small,$^{34}$ we do not only analyze the defect
free \dwp\ but also the defect related DWP.

To define the coordination of a silicon atom in a classical pair
potential it is necessary
  to introduce maximum bond cutoffs to determine if a bond exists or not. The minimum between
  the first and the second nn-shell in the
  radial distribution functions for the minimized structures lies around $2.2\ang$. This value
   has thus been used as a cutoff to define bonds, Si--O distances below this value are
  considered bonds and larger distances are not. The general results do not critically
  depend on the used cutoff value as a broad minimum region between the two nearest-neighbor-shells
  exists.

\subsection{Lithium silicate}
For the alkali silica simulations a 153 particle system with only 2
Li-atoms has been used. The low lithium concentration guarantees,
that the Li-atoms behave almost independently and that the network
structure is rather similar to that of pure silica. The potential
has been taken from Habasaki and Okada.$^{35}$ Its alkali-free limit
is close to the BKS-potential. The simulations have been conducted
under the same conditions as for pure silica.

\subsection{Search for DWP}

The key idea of our search algorithm is to start from one
minimized configuration, i.e. a local minimum of the PEL, and then
to perform a specified number of MD steps. Afterwards the system
is minimized again. When this procedure ends in a configuration
which is different from the starting configuration it is checked
whether there exists a saddle point between both configurations.
 If yes, one has identified
one DWP. This procedure is repeated many times (typically: 100
times) in order to identify most if not all DWP. A DWP is
characterized by three parameters: its asymmetry $\Delta$, its
potential height $V$ and its distance between both minima $d$. More
specifically we either use the Euclidean distance $d$ or the
mass-weighted distance $d_{mwrp}$ along the reaction path.$^{16}$
Since these differences are not relevant in the context of the
present work they will be neglected in what follows. For the
determination of the saddle a robust saddle search routine has been
chosen.$^{31}$

A DWP is kept for the further analysis if $\Delta < \Delta_{max}$
and $d < d_{max}$ where $\Delta_{max}$ and $d_{max}$ are specified
values. This limitation is essential to enable a systematic search.
DWP with, e.g.,  very large values of the asymmetry are difficult to
find such that a systematic search of {\it all} DWP (starting from a
given configuration) is not possible within reasonable computer
time. In contrast, if one is only interested in DWP within the
range, specified above, one may hope to find (nearly) all DWP. As
will be shown below, the TLS are confined to this range, i.e. TLS
with $d > d_{max}$ are irrelevant.

A simple way to estimate the degree of completeness of the search
procedure is to analyze how often a specific DWP is found during the
repeated simulations. It has been shown that for appropriately
chosen parameters a systematic search for DWP is indeed possible for
the BMLJ system $^{16}$ as well as for silica $^{25}$ and properties
of TLS, i.e. of nearly symmetric DWP, can be extracted. For a
successful assessment of TLS properties one first has to deal with
the strong positive correlations among all three parameters $d$,
$V$, and $\Delta$. Guided by the soft-potential model $^{36,37}$ we
have mapped the distribution $p(d,V,\Delta)$ on a distribution of
soft-potential parameters $w_2,w_3,w_4$ which fortunately to a very
approximation turn out to be independent of each other. Based on the
independent distribution functions $p_i(w_i)$ it is possible to
generate a sufficiently large set of DWP with the same statistical
properties as the initial set of DWP. In this way one can cover the
full parameter range with high precision and in particular obtain
the distribution of TLS as a subset of the whole parameter range. A
closer description of this {\it parametrization method} as well as
further technical details can be found in Refs.$^{16,17}$.

The parameters, used for our simulations, are given in Tab.1. For
the characterization of the DWP we first determine which particle
moves most during the transition. This particle is denoted {\it
central} particle. For BMLJ we have therefore distinguished A-type
DWP where the central particle is an A-particle and, analogously,
B-type DWP. For silica we have defined \dwpnd\ as the subset of DWP
where the initial and the final configuration are defect-free and no
bond breaking occurs during the transition. The remaining set is
denoted \dwpod. For lithium silicate we define \dwpli\ as those DWP
for which one of the two lithium atoms is the central particle. The
number of DWP, found in our simulations, are also listed in Tab.1.

\section{\label{results}RESULTS}

\subsection{Properties of DWP parameters}

 \begin{table}
    \begin{tabular}{|l||c|c|c|}
          \hline
       & \# DWP  & $d_{max}$& $\Delta_{max}$
        \tabularnewline \hline
      BMLJ (A-type)  & 521 & 0.8 & 0.5 \\
       BMLJ (B-type) & 6001 & 0.8 & 0.5 \\
      \dwpnd\    & 250 & 5 \AA & 1500 K \\
      \dwpod\     & 1419 & " & " \\
      \dwpli\    & 1885  & " & " \\ \hline
    \end{tabular}
  %\end{minipage}
  \vspace*{0.5mm}
  \caption{
  \label{tab1}The number of DWP analyzed for all different systems (system sizes: N=65 for BMLJ; N=150 for silica; N=153
  for lithium silica). They were recorded if
   their values of $d$ and $\Delta$ satisfy $d < d_{max}$ and $\Delta < \Delta_{max}$, respectively.}
  \end{table}

\begin{figure}
    \begin{center}
    \includegraphics[width=14cm]{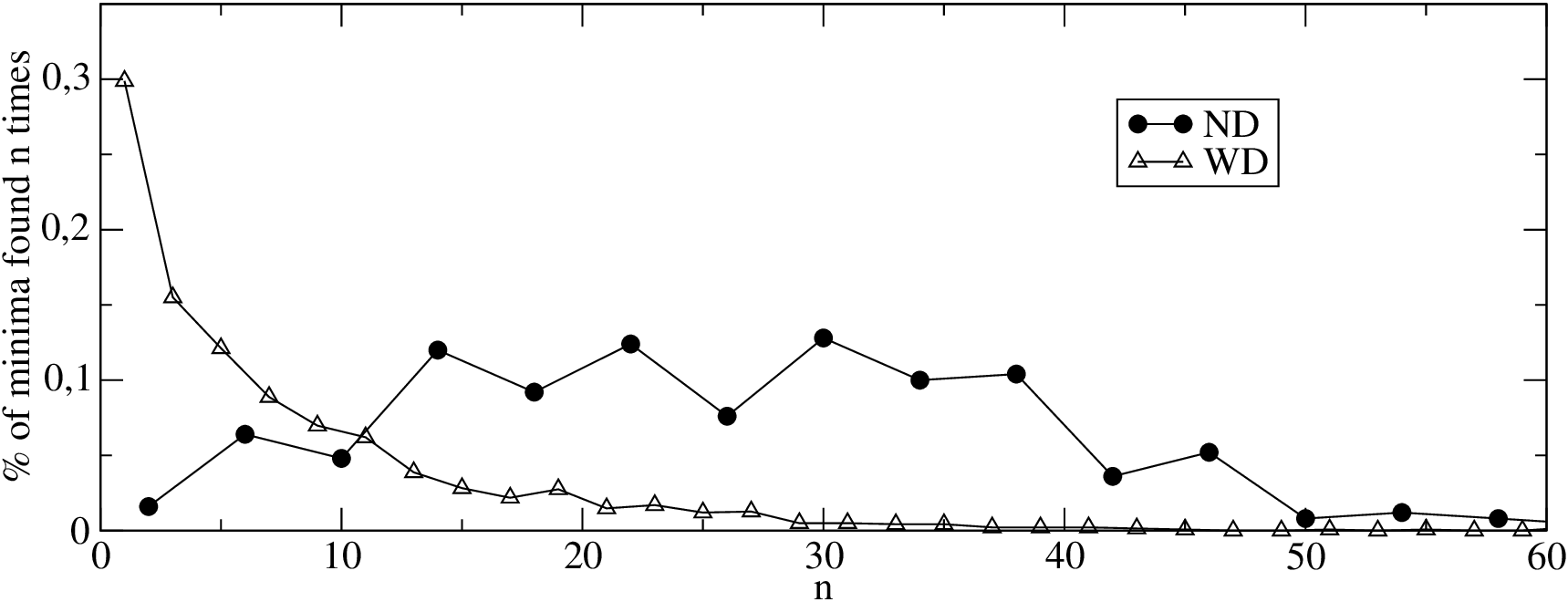}
    \caption
    {  \label{SiO2_completeness}The histogram of how often a nearby minimum was found per starting minimum, using
    100 independent runs.
        }
    \end{center}
  \end{figure}

Before analyzing the total number of DWP, found in our simulations
with the choice of $\Delta_{max}$ and $d_{max}$ as listed in
Tab.\ref{tab1}, one has to ask whether or not the search for DWP was
complete. In Fig.\ref{SiO2_completeness} we display the results for
\dwpnd\ and \dwpod\ when applying the completeness criterion. The
curve for \dwpnd\ behaves similarly to what we have found for the
BMLJ system.$^{16}$ It displays a maximum for a value much larger
than one (here: approx. 30). Qualitatively, this means that whenever
a DWP is present it is found quite often during the repeated runs.
As a consequence it is statistically very unlikely that a DWP is not
found at all. This is exactly the condition for a (nearly) complete
search. This result has been already used in our previously
published work.$^{25}$ In contrast, the probability that a \dwpod\
is found just once out of the 100 attempts is significantly larger
than to be found twice or even more often. Thus, it is likely that
several \dwpod\ are not found at all. A similar observation is made
for the \dwpli\ (data not shown). It is, however, unlikely that the
true number of DWP is orders of magnitude larger than the number of
DWP already found because otherwise the probability to find a DWP
only once or twice would be much larger than 30\% for \dwpod.
Furthermore for the estimation of the TLS one mainly needs to have
reliable information about DWP which are nearly symmetric and have a
small value of $d$ (see below for more details). It is likely that
this subset of DWP is found with a higher probability because of its
proximity on the PEL which further supports that at least the number
of TLS, estimated below, is not far away from the true value.

Based on the number of DWP, found in our simulations and listed in
Tab.\ref{tab1}, and the number of analyzed initial configurations
one can estimate the probability that a given configuration contains
one DWP within the specified parameter range. For atomic systems
like BMLJ it is useful to divide the number of DWP also by the
number of particles per configuration to obtain a per-particle DWP
probability. Thus, for the A-Type DWP we have defined the
probability relative to the number of A-particles. The B-type DWP
are treated analogously. For the lithium silicate system we consider
the occurrence of \dwpli\ relative to the two individual lithium
atoms per configuration. For pure silica a SiO$_4$-tetrahedron
serves as an elementary unit because of its rigid character at low
temperatures. Thus a silica configuration with $N=150$ particle
contains 50 individual elementary units. Actually, in some models of
silica the tetrahedra are directly treated as rigid bodies.$^{18}$
We note in passing that the size of the elementary unit can be
estimated from analyzing the material constants of silica.$^{38}$
Therefore, for a better comparison with other systems the number of
\dwpnd\ for defect-free configurations is related to the number of
tetrahedra. It turns out that for a silica configuration with
$N=150$ atoms the probability to find a DWP is 10 times larger if
the configuration contains a defect. Thus the DWP can be to a good
approximation exclusively related to the presence of a defect. On
average, it turns out that for $N=150$ a configuration contains
approx. 1.2 independent defects. Using all these pieces of
information one can express the number of \dwpod\ relative to the
number of elementary units, which here are the number of independent
defects. For all five different types of DWP the probability of DWP
formation per elementary unit (within the specified range of
parameters) is given in Tab.\ref{tabnr}. Note that for \dwpod\ and
\dwpli\ we can only determine lower limits.

   \begin{table}
    \begin{tabular}{|l||c|c|c|}
      \hline
         & DWP/elem.unit & $f_\Delta$ & TLS/elem.unit  \\
   \hline
      BMLJ (A-type)& $1.0 \cdot 10^{-3}$ & 2.6 &  $1.1 \cdot 10^{-5}$ \\
      BMLJ (B-type) & $4.0 \cdot 10^{-2}$ & 2.6& $5 \cdot 10^{-4}$ \\
      \dwpnd\     & $1.4 \cdot 10^{-3}$ & 2.0 & $3.8 \cdot 10^{-6}$  \\
      \dwpod\       & $\ge 5 \cdot 10^{-1}$ & 1.1 & $\ge$ $7.3 \cdot 10^{-4}$  \\
      \dwpli\     & $\ge 2.8 \cdot 10^{-1}$ & 1.25 & $\ge 4.7 \cdot 10^{-4}$
      \\\hline
    \end{tabular}
  %\end{minipage}
  \vspace*{0.5mm}
  \caption{
 \label{tabnr}The probability that a specific elementary unit is the central
      unit for the transition of a DWP with asymmetry less than 2K, i.e. a TLS.
      $f_\Delta$ characterizes the deviations from a constant distribution of asymmetries. }
  \end{table}

  \begin{figure}
    \includegraphics[width=14cm]{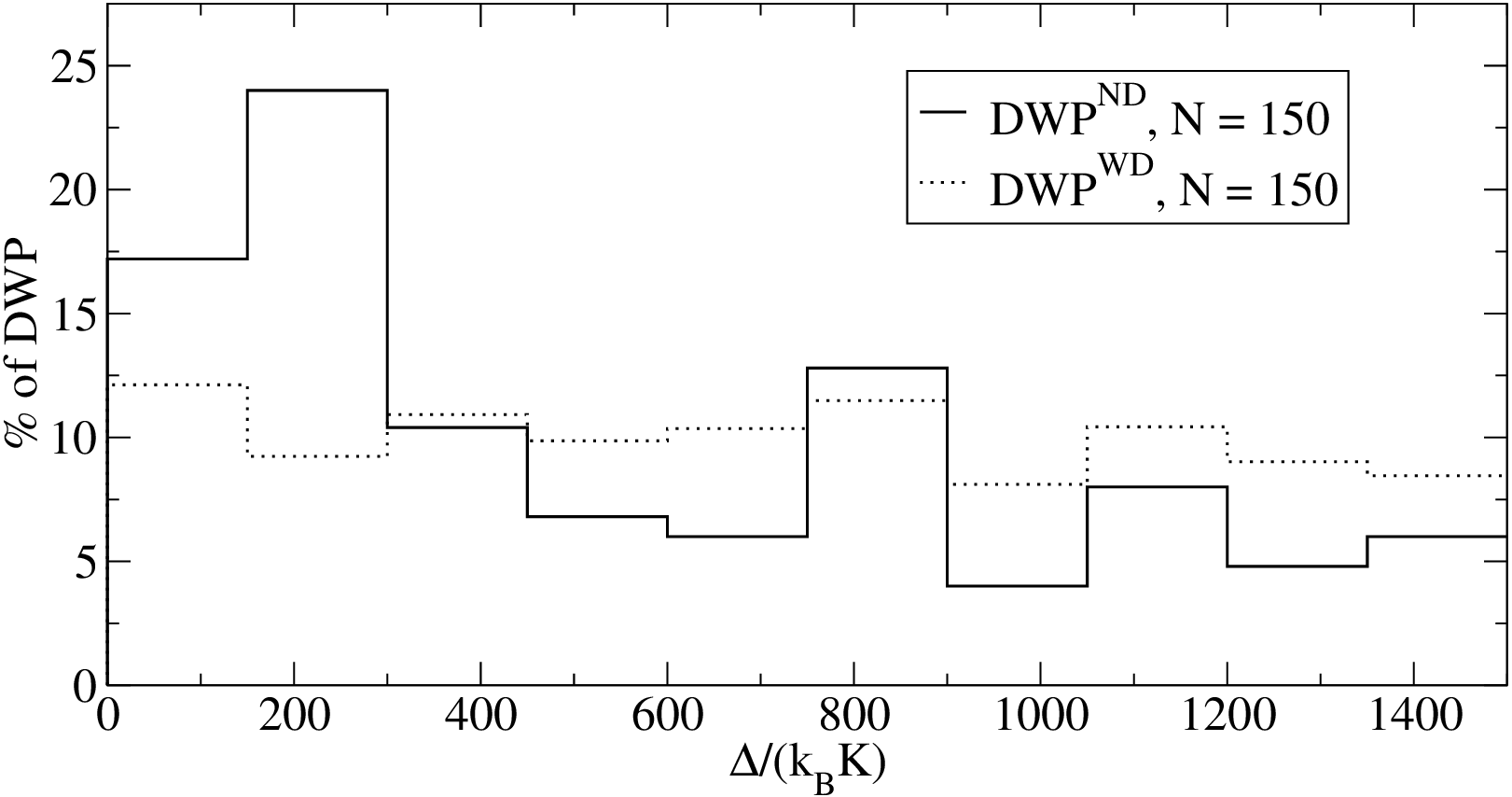}
   \caption{
  \label{SiO2_150_nd_od_asym_distr}The distribution of asymmetries for the \dwpnd\ and \dwpod.}
  \end{figure}

\begin{figure}
  \includegraphics[width=14cm]{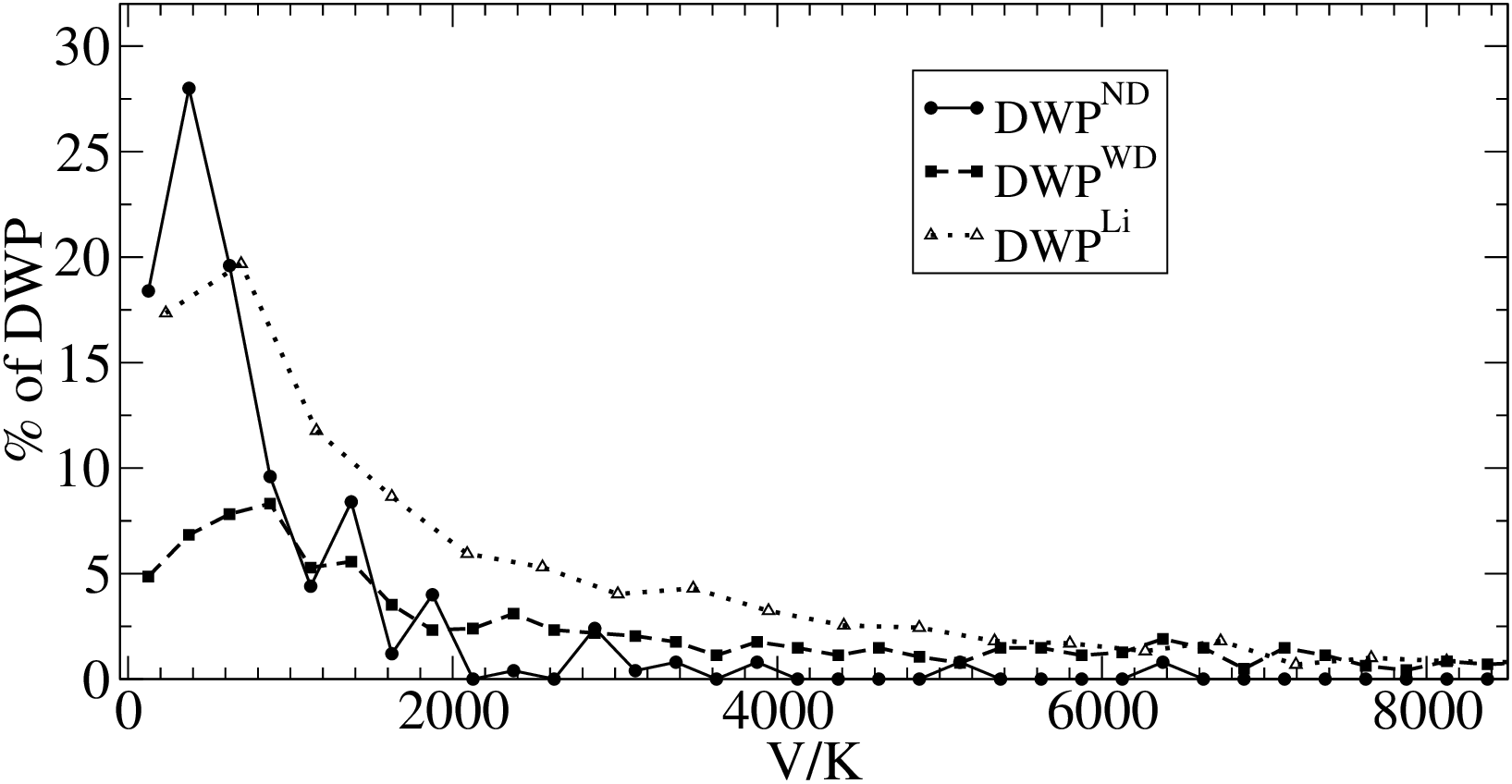}
  \caption{\label{all_barr_distr} Barrier distribution for the different types of DWP.}
\end{figure}

For understanding the low-temperature anomalies one is interested in
the number of TLS (here defined via $\Delta < 2$ K) rather than in
the number of DWP. In analogy to the DWP we define the \tlsnd,
\tlsod\ and \tlsli. If the distribution of asymmetries were constant
in the interval $[0,\Delta_{max}]$ the number of DWP with an
asymmetry smaller 2K could be directly calculated by multiplying the
above values with 2K$/\Delta_{max}$. Closer inspection of the
distribution shows, however, that the density is somewhat higher for
very small asymmetries; see Fig.\ref{SiO2_150_nd_od_asym_distr} for
\dwpnd\ and \dwpod. The density for $\Delta < 300$ K is roughly
constant within statistical noise. One can define a factor
$f_\Delta$ which describes the increases of the density in the limit
of low asymmetries as compared to the prediction for constant
density in the whole interval. An estimation for this value is also
shown in Tab.\ref{tabnr}. As can be judged from
Fig.\ref{SiO2_150_nd_od_asym_distr} there is some statistical
uncertainty of approx. 20\%.  The value of $f_\Delta$ is
particularly large for \dwpnd\ and the DWP in the BMLJ system.
Qualitatively, it is related to the barrier distribution in
Fig.\ref{all_barr_distr}. Due to the strong correlation of asymmetry
and barrier height DWP very high barrier heights are very unlikely
to have a small asymmetry. Thus a broad distribution of barrier
heights implies that only a smaller fraction of DWP can be relevant
for the low-temperature anomalies. Indeed, it turns out that to a
good approximation the value of $f_\Delta$ is proportional to the
fraction of DWP with $\Delta < 1000$ K in Fig.\ref{all_barr_distr}.
Now the number of TLS per elementary unit can be directly estimated.
The results are also given in Tab.\ref{tabnr}.

As a main conclusion from this analysis it turns out that A-type DWP
for the BMLJ system and \dwpnd\  are very rare. The other extreme
are the extrinsic lithium defects \dwpli\ as well as the intrinsic
defects \dwpod\ in silica. Thus the presence of a small particle in
a disordered network or the breaking of this network, e.g. via
non-bridging oxygens (see below), prompts the formation of bistable
modes, i.e. DWP. Hence via local motion of the defect the system can
acquire a new metastable position. Also the probability of the
formation of B-type DWP for BMLJ is dramatically enhanced. This is
consistent with the view that the B-particles can be regarded as
defects among the majority of the A-particles. We note in passing
that for the lithium silicate system also the number of DWP with
silicon or oxygen as central particles is increased by approx. 50\%
if compared to the number of DWP for an average silica
configuration. This may be related to the fact that the presence of
lithium automatically implies a breaking of the silicate network and
then the additional intrinsic defects may give rise to an increased
number of DWP.

A central DWP parameter is the barrier height. In
Fig.\ref{all_barr_distr} we display the distribution of barriers for
the different types of DWP. Interestingly, the absolute values of
the barrier heights are very different. For \dwpnd\ the transition
between both minima can be described as a coupled rotation of
several SiO$_4$-tetrahedra.$^{25}$ Obviously, this can be achieved
by surmounting only a relatively small energy barrier. A
rationalization for the higher barriers for \dwpod\ and \dwpli\ will
be presented further below. In Ref.$^{39}$ the distribution for
\dwpnd\ has been estimated to be confined to approx. $V < 1000$ K.
This is in very good agreement with the data shown in
Fig.\ref{all_barr_distr}.

\begin{figure}
  \includegraphics[width=14cm]{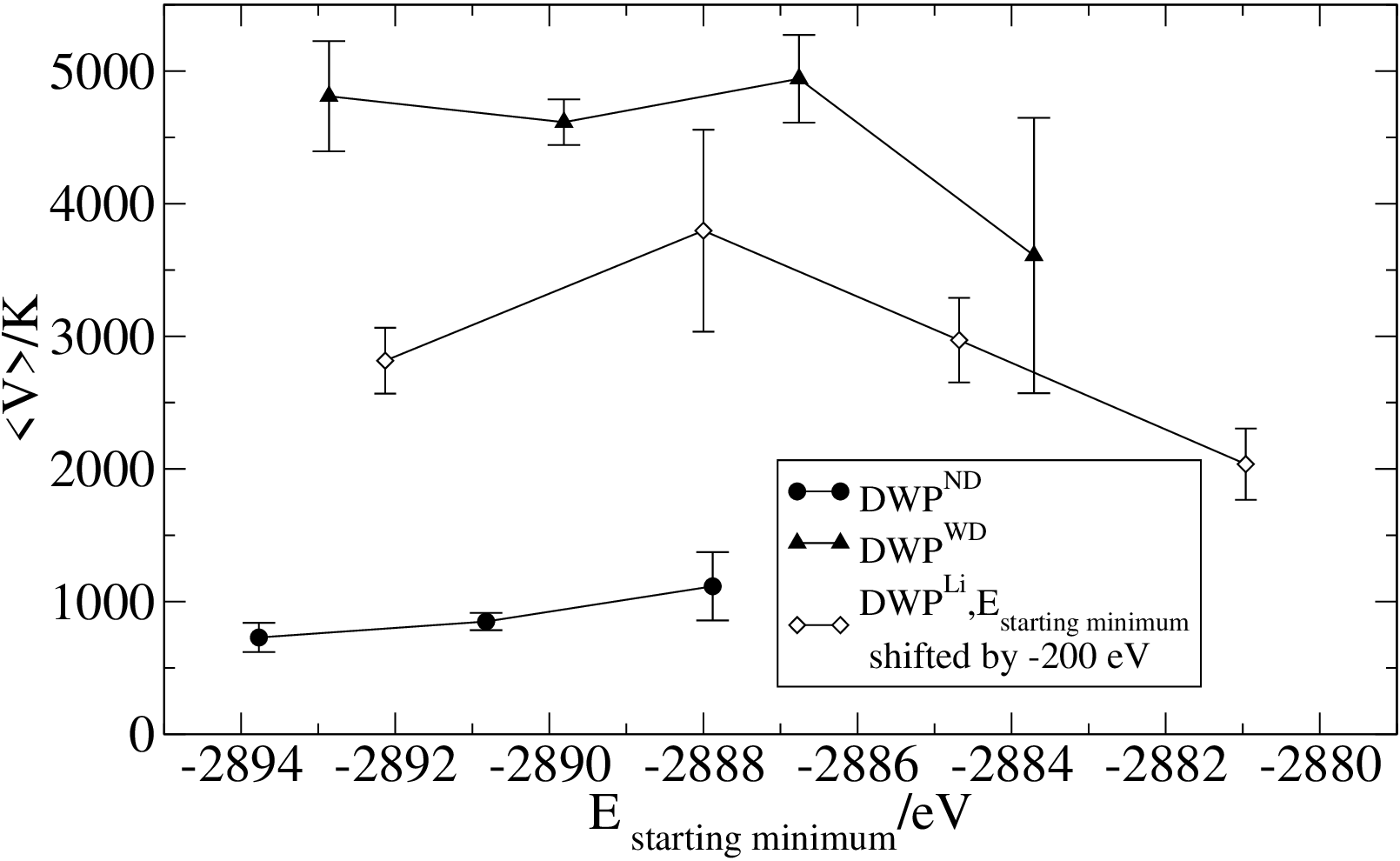}
  \caption{
  \label{all_V_E_corr}Dependence of the average DWP barrier on the energy of the initial configuration.
     The \dwpli\ data are shifted by -200 eV to handle the
           different minimum energies.}
\end{figure}

Furthermore, one may ask whether the properties of the DWP depend on
the energy of the starting configuration. Therefore we analyze the
average barrier height of the DWP in dependence of this initial
energy; see Fig.\ref{all_V_E_corr}. The energies correspond to the
relevant energies at low temperatures. Actually, the previous
analysis about the relaxation properties of silica has revealed that
the PEL of silica has a low-energy cutoff which for the present
system size can be estimated to be around -2895 eV.$^{40}$ For all
three systems there is no significant dependence of the average
potential height on energy relative to the level of statistical
uncertainties. Thus, one may conclude in analogy to our previous
results for the BMLJ system, that the fast cooling rate in computer
simulations is no serious problem for the quantitative analysis of
DWP. Also different quantities, analyzed along this line, do not
show a significant energy-dependence.

\begin{figure}
  \includegraphics[width=14cm]{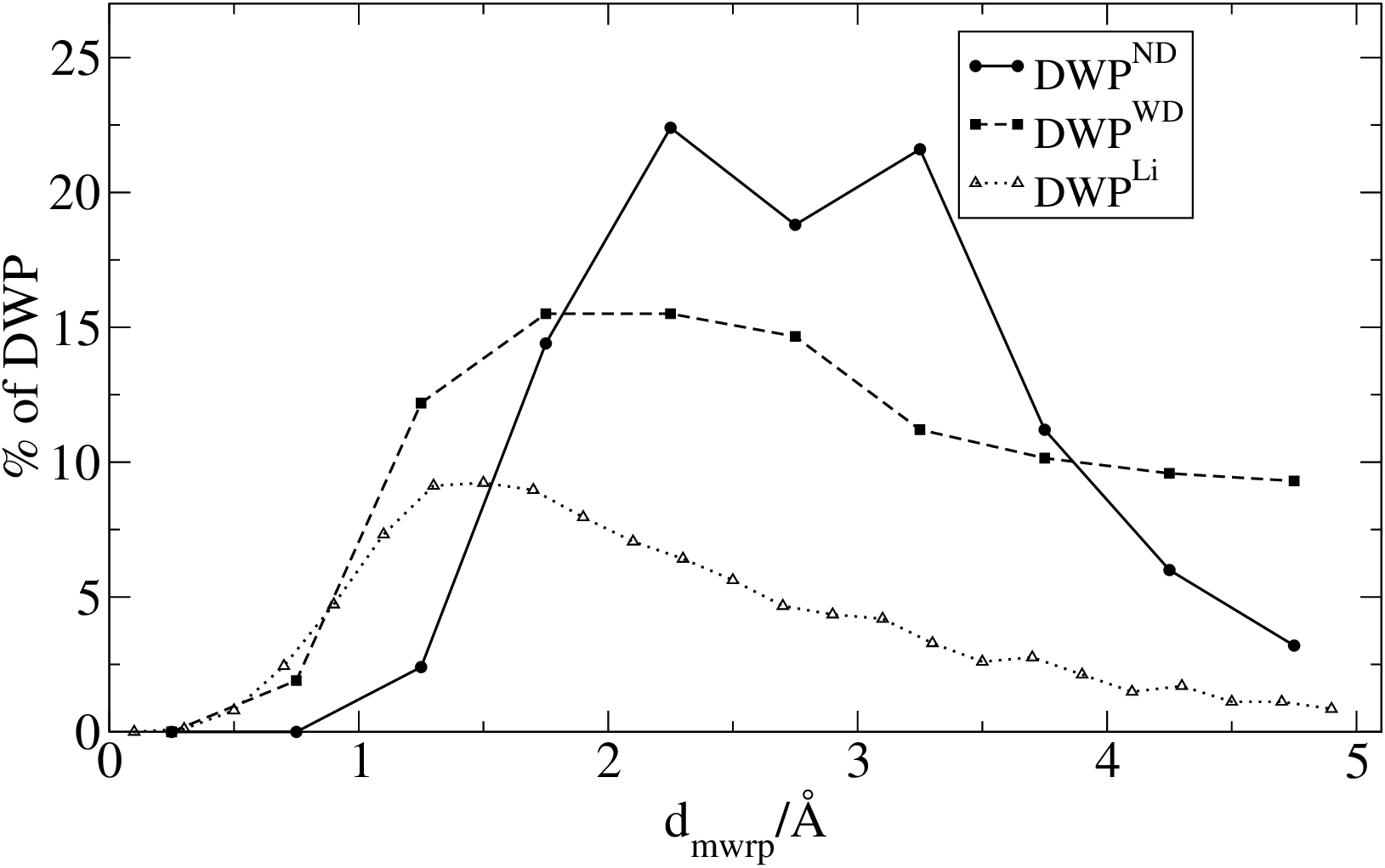}
  \caption{
\label{all_d_distr}The distribution of distances for different DWP.}
\end{figure}

Finally, we have analyzed the distance $d$ between the two minima of
the DWP. The results are shown in Fig.\ref{all_d_distr}. Nearly all
\dwpnd\ display distances between 1.5 \AA\ and 4 \AA. This implies
that the motional pattern of coupled rotations of tetrahedra
requires, on the one hand, some minimum length and, on the other
hand, does not extend beyond some upper limit. In contrast, in
particular for the defect dynamics, i.e. \dwpod\, a larger variance
of possible distances is observed. This hints towards a larger
number of different motional mechanisms in the presence of defects.

\subsection{Microscopic properties of DWP}

A central question deals with the number of atoms involved in the
transition. This can be captured by the participation ratio. In
literature different definitions can be found. Here we use $R^H$
and $R^S$ as defined in Tab.\ref{tabpart}.  d$_{max}$ is the
distance moved by the central particle and $d_i$ the distance
moved by the i-th particle. The results are listed in
Tab.\ref{tabpart}.

   \begin{table}
  %\begin{minipage}[t]{\textwidth}
    \begin{tabular}{|p{3.5cm}||
                    c|
                    c|
                    c|} \hline
                               &
      \multicolumn{1}{c|}{$R^H$}   &
      \multicolumn{1}{c|}{$R^S$}           \tabularnewline
                               &
      \multicolumn{1}{c|}{$\langle d^2/d_{\text{max}}^2\rangle$}   &
      \multicolumn{1}{c|}{$d^4/\sum_i d_i^4$} \\
      \hline
      BMLJ ($A$-type)         & 6.7 & 16.1     \tabularnewline[1mm]
      BMLJ ($B$-type)         & 3.0 & 9.0      \tabularnewline[1mm]
      \dwpnd, $N=150$   & 9.1  & 24        \tabularnewline[1mm]
      \dwpnd, $N=600$   & 9.0  & 31        \tabularnewline[1mm]
      \dwpod, $N=150$   & 6.4  & 19       \tabularnewline[1mm]
      \dwpli, $N=153$   & 1.9  & 3.6       \\ \hline
    \end{tabular}
  %\end{minipage}
  \vspace*{0.5mm}
  \caption{
 \label{tabpart}Participation ratios for the different systems studied in this work.}
  \end{table}

As the most prominent feature the transition in \dwpli\ is extremely
localized. Thus one may speak of a single-particle transition. This
agrees with our previous observation for the BMLJ system. There the
B-type DWP, corresponding to the dynamics of the small B-particles,
are localized, too.  In contrast, larger participation ratios can be
found for the A-type DWP in the BMLJ system as well as the DWP in
the silica system. Similar values have been obtained in
Refs.$^{19,41}$. Taking into account that for silica the elementary
units are SiO$_4$ tetrahedra and not individual atoms, these values
should be considered as upper limits for the number of independent
degrees of freedom, involved in the transitions. For \dwpnd\ we have
performed a systematic search for $N=150$ and $N=600$ particles. In
agreement with other observables finite size effects are only weak.
Thus, one may conclude that neither the finite size of our sample
nor the fast quenching rate, used in computer simulations, have a
relevant effect on the properties of the DWP.

For the case of the BMLJ system the participation ratio is slightly
lower for DWP with smaller distances $d$ (unpublished work). Since
TLS have smaller distances than the average DWP, the above values
for the participation ratio should be regarded as upper limits for
TLS. Thus, one may conclude that in all cases one has less than,
let`s say, 10 particles which participate in a tunneling mode.

\begin{figure}
  \includegraphics[width=14cm]{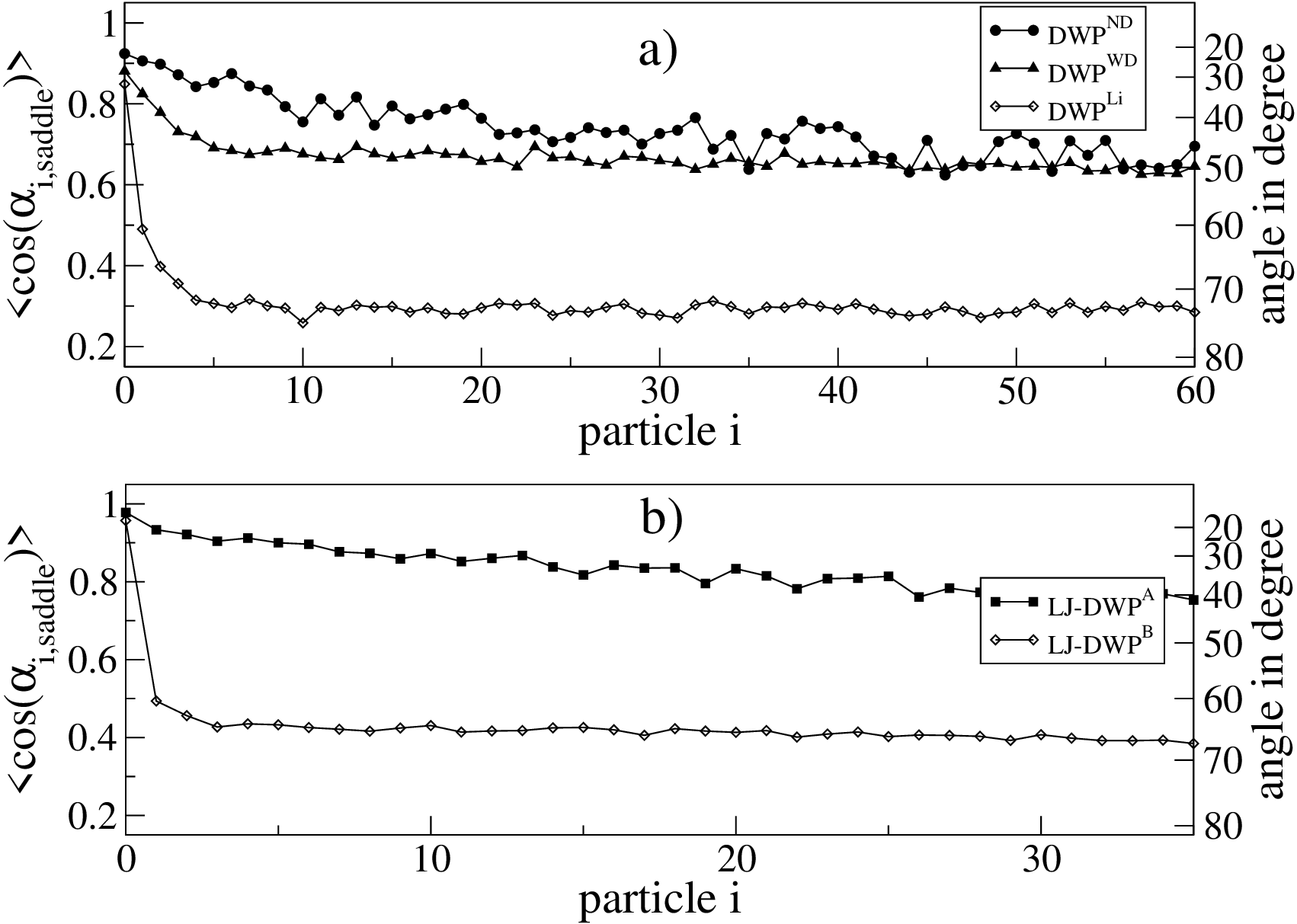}
  \caption{
 \label{all_saddle_angle}The average cosine of the angle between the transitions from the first minimum to
  the saddle and the transition from saddle to the second minimum. The particles are sorted
 according to their absolute value of the Euclidean distance when moving between both minimum configurations.}
 \end{figure}

For a closer analysis of the nature of the transitions we have
analyzed whether the transition between both minima for the
different particles is basically a straight line in real space or
whether it is strongly curved. For this study we compare for all
particles the transition vector from the first minimum to the saddle
with the corresponding vector from the saddle to the second minimum
and determine the angle $\alpha$ between both vectors. In case of a
straight line both vectors are parallel, i.e. $\alpha = 0$. In the
other extreme case a particle performs a forward-backward motion
such that both vectors are antiparallel, i.e. $\alpha = \pi$.
Furthermore, we sort the particles according to the Euclidean
distance between the initial and final position. The particle with
index $i=0$ thus corresponds to the central particle, the particle
with index $i=1$ to the particles moving second most and so on. This
allows us to average over all DWP. The results are shown in
Fig.\ref{all_saddle_angle}.

Clearly, for all systems the central particle of a DWP moves along a
nearly straight line during the transition ($\langle \cos \alpha
\rangle \approx 0.9$). However, for the other particles major
differences exist. For A-type DWP of the BMLJ system all particles
perform a relatively straight transition path. This reflects the
cooperative nature of the transition. In contrast, for B-type DWP
already the transition path for the particle with $i=1$ displays a
strong curvature. This is compatible with the single-particle nature
of the DWP. During the transition of a small B-particle between two
local minima of the PEL the surrounding A-particles mainly retreat
during the transition thereby reducing the value of the saddle
energy and finally go back to a similar position as before.

Not surprisingly, the same difference is observed when comparing
\dwpnd\ and \dwpod\ with \dwpli. The one-particle type transitions
for \dwpli\ are, in analogy to the B-type DWP, connected with a
strong curvature for the transition of the remaining particles.
There are, however, additional differences between \dwpnd\ and
\dwpod\ for small particle index $i$. The coupled
tetrahedra-rotation for \dwpnd\ implies a similar behavior of the
most mobile particles. In contrast, for the transitions involving
defects the degree of cooperativity is somewhat smaller.

Comparing \dwpnd\ with the A-type DWP from the BMLJ system it
turns out that that the angles are generally larger in the silica
system. The motion is thus more curved, which seems reasonable due
to the network structure and fits to displacements due to rotating
tetrahedra.

\subsection{Microscopic mechanism of the DWP transition}

In Ref.$^{25}$ it has been shown that for \dwpnd\ the crucial step
during the transition is a Si-O-Si bond flip. Correspondingly, in
more than 99\% of all cases the central particle is an oxygen and
only starting from particle index $i=10$ (see above for its
definition) the probability to find a silicon atom is close to the
statistical value of $1/3$. Qualitatively, the dominance of oxygen
atoms as the most displaced particle is compatible with the picture
of cooperative tetrahedra rotations which mainly involve the
dynamics of oxygen atoms. Interestingly, it turns out that oxygen
atoms, acting as central atoms, possess a significantly shorter Si-O
bond length. This structural motif seems to enhance the probability
for the formation of a DWP.$^{25}$

No simple picture for \dwpod\ can be formulated. Rather we have
observed a variety of different mechanisms from a close inspection
of 30 DWP. Most of the times the transition could be described by
one of the following three mechanisms: (1) A bond between a silicon
and a threefold coordinated oxygen breaks, the silicon forms a new
bond with another oxygen. This oxygen is now threefold coordinated.
(2) A bond with a threefold coordinated oxygen breaks and a silicon
center changes its coordination from four to three or from five to
four. (3) A dangling oxygen (coordinated to only one silicon) forms
a bond to a three- or fourfold coordinated silicon resulting in a
four- or fivefold coordinated silicon and sometimes another oxygen
from that silicon becomes either a dangling oxygen itself or
coordinates to another silicon. The variety of these scenarios in
part also reflects the different types of DWP. Not surprisingly, in
this case the probability that the central particle is an oxygen
atom is only slightly enhanced as compared to the statistical value
(75\%).

 \begin{figure}
  \includegraphics[width=14cm]{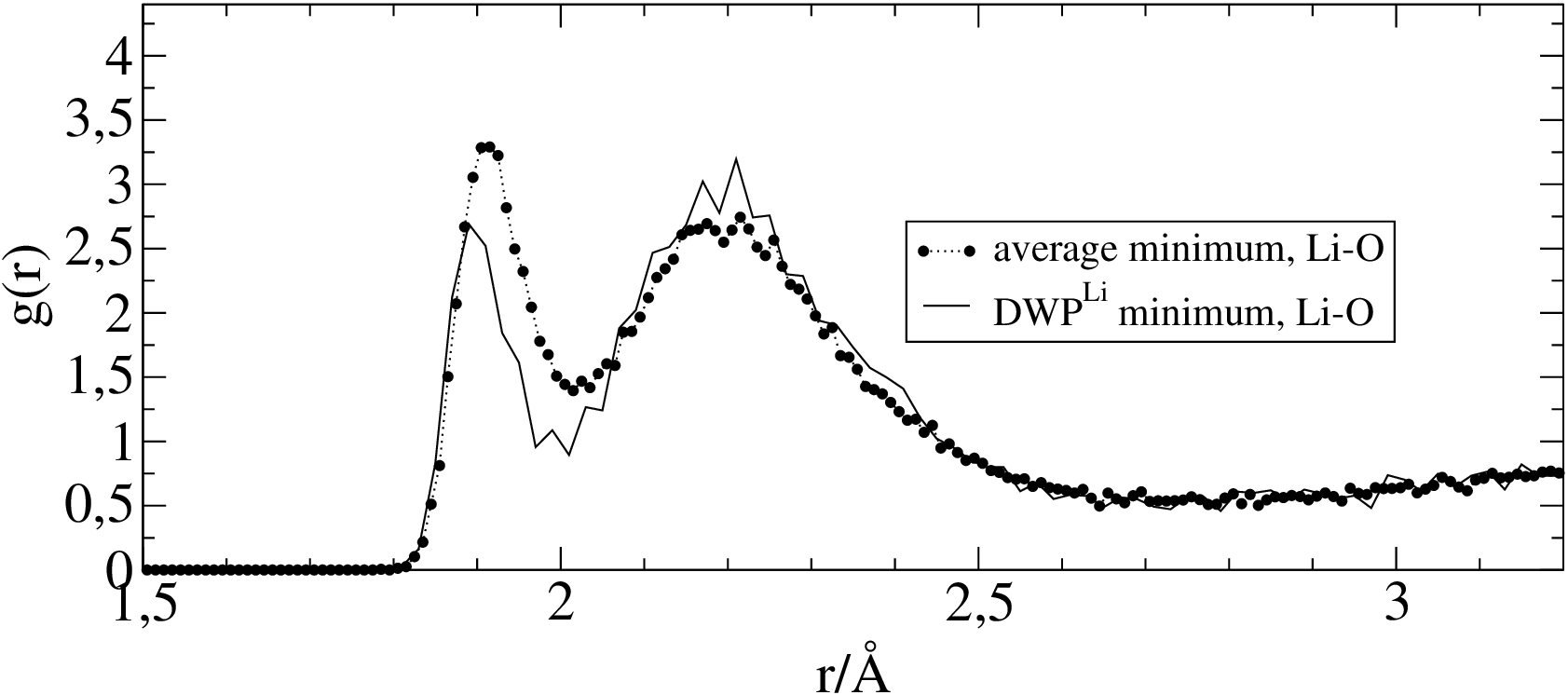}
  \caption{
  \label{sio2li_gofr_3_8}Partial pair correlation function $g_{Li-O}(r)$
    for the central lithium of the \dwpli\
    in comparison
    with the average radial distribution function. Only the first peak is shown as
    no differences exist in other sections.
    }
 \end{figure}

To study the case of \dwpli\ we have analyzed the partial Li-O pair
correlation function for the minimum configuration. The results for
the nearest-neighbor shell are shown in Fig.\ref{sio2li_gofr_3_8}.
The first peak of this double-peak structure is mainly related to
non-bridging oxygen the second peak to bridging oxygen. The density
of oxygen atoms, related to the first peak (and thus being mainly
non-bridging oxygen), is decreased by approx. 25\% when a DWP is
present. The integration of both peaks shows that density is
transferred from the first to the second peak, but no density is
lost from the first two peaks. Thus a reduction of the density of
non-bridging oxygen and a corresponding increase of the number of
bridging oxygen is a prerequisite for the formation of a DWP.
Interestingly, the average number of oxygen atoms in the
nearest-neighbor shell decreases from 3.9 to 3.4 when comparing the
minimum with the saddle. Thus, during the transition the lithium ion
changes part of its oxygen neighborhood.

\subsection{Dipole Moment}

\begin{figure}
  \includegraphics[width=14cm]{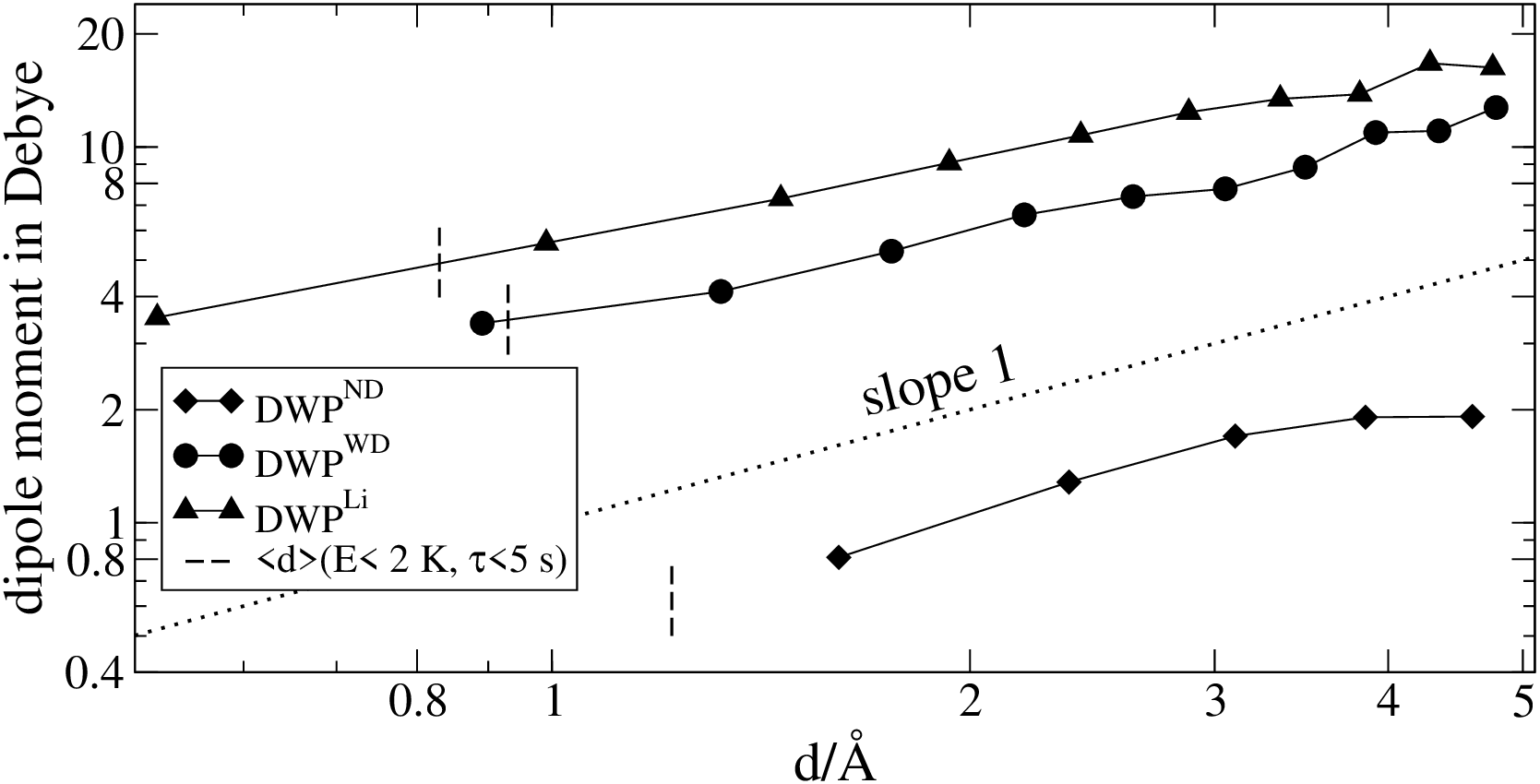}
  \caption{
 \label{dipole_vs_d}The dipole moment for \dwpnd\, \dwpod\ and \dwpli\ in dependence of the distance
     between both minima. Included is a line of slope 1 and the average distances, estimated for the TLS.
     }
 \end{figure}

Of major experimental interest is the coupling of the TLS to
electric fields via their dipole moment. It can be easily
determined via
\begin{equation}
\label{eqdip} \vec{M} = \sum_i q_i \vec{d_i}
\end{equation}
where $q_i$ denotes the partial charge and $\vec{d}_i$ the
translational vector of particle $i$ during the transition. Of
interest is the absolute value of the dipole moment $M =
\sqrt{|\vec{M}|^2}$ which can be calculated in a straightforward
manner from Eq.\ref{eqdip}. More precisely, we have averaged $M^2$
and finally calculated the square root. The partial charges in our
potentials are 0.87e for lithium, 2.4e for silicon and -1.2e for
oxygen. Note that a single-particle transition with $d = 1$\AA\ and
unit charge would correspond to 4.8 Debye. The results are shown in
Fig.\ref{dipole_vs_d} for the different types of DWP. We have
averaged DWP with similar distances $d$, in order to keep track of a
possible dependence on $d$. To a good approximation the data can be
described by a linear relation, i.e.
\begin{equation}
\label{eqmallg}
 M = \zeta q d
\end{equation}
where $q$ is the average partial charge and $\zeta$ a dimensionless
proportionality constant. In order to convey a feeling for the
absolute value of $M$ it may be instructive to estimate $M$ for the
simple scenario for which all particles move independently, i.e. the
$\vec{d}_i$ are uncorrelated. In this limit one obtains the same
expression as Eq.\ref{eqmallg} with $\zeta = 1$. Thus the
proportionality to $d$ is a generic property which results from the
definition of the dipole moment whereas the value of $\zeta$
contains information about the motional mechanism.

For \dwpli\ one would expect $M(2\ang) \approx 8$D (using the
partial charge of lithium) which is close to the numerically found
value. Thus we find $\zeta \approx 1$. In any event, for a localized
transition any correlation effects between adjacent particles cannot
be relevant for the dipole moment. The other extreme case is \dwpnd\
for which the dipole moment is approx. 10 times smaller than
expected for the uncorrelated case. This clearly shows that the
dynamics is highly cooperative. In general, two effects may give
rise to the reduction. First, particles of opposite charge move in
the same direction; second, identical particles move in opposite
directions. The latter case is relevant for \dwpnd\ because of the
dominance of tetrahedra rotations, involving significant oxygen
motion. In case of ideal rotations around a threefold axis the
resulting dipole moment would be zero. Due to the non-symmetric
nature of the actual rotational axis some residual dipole moment
remains.$^{25}$

For \dwpod\ the reduction as compared to the statistical value is
only a factor of approx. 3 (taking into account the somewhat higher
effective charge because also silicon atoms can move significantly).
Despite the similarity in the participation ratios of \dwpnd\ and
\dwpod\ this shows that the nature of the cooperativity is
significantly different. Beyond the rotation of tetrahedra it is the
transfer of the defect structure which plays an essential role and
which does not possess the cancelation effect of pure tetrahedral
rotations.

\begin{figure}
  \includegraphics[width=14cm]{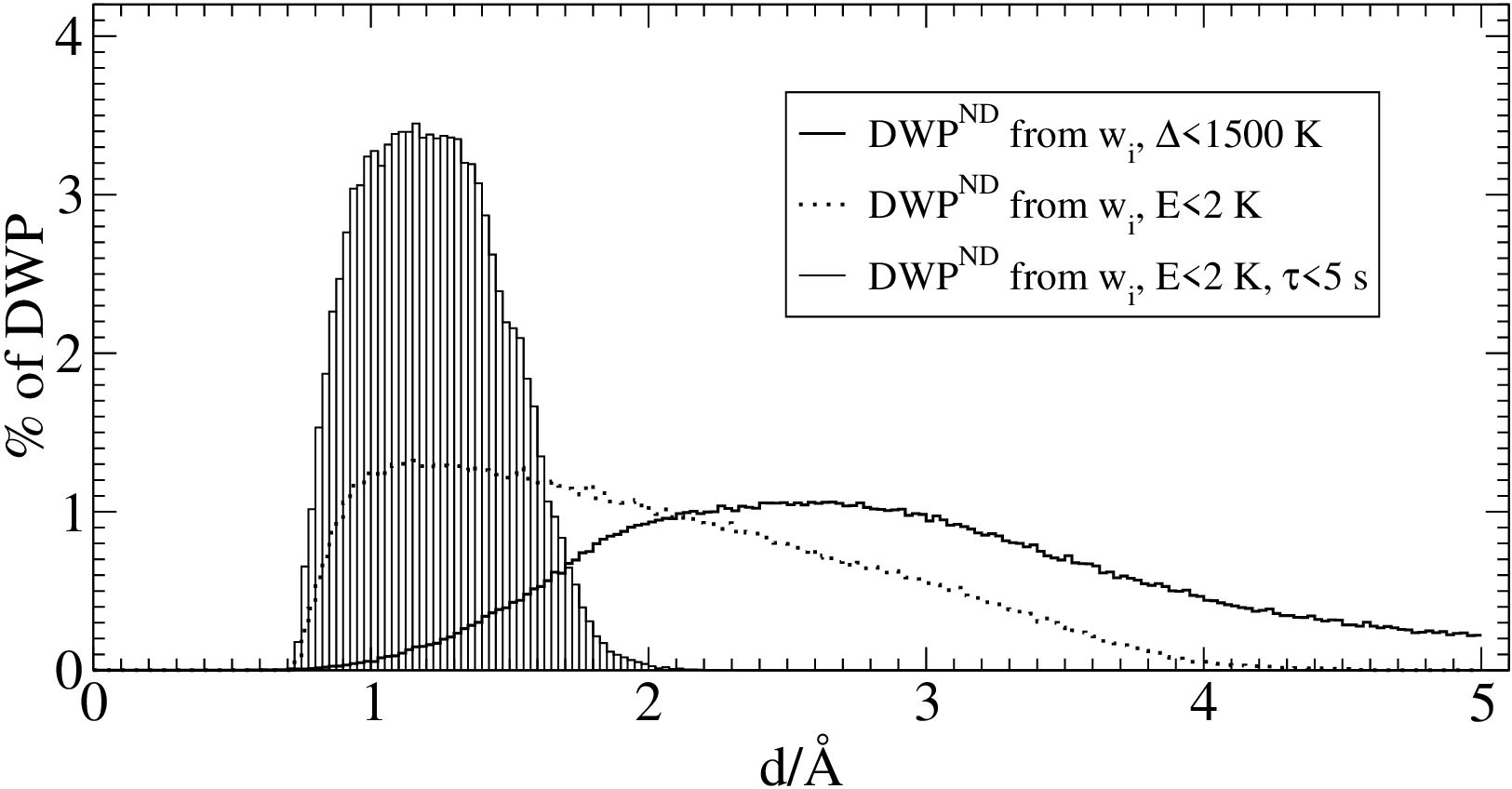}
  \caption{
 \label{SiO2_d_distr_various}The distance distribution for a subset of \dwpnd\, fulfilling different criteria
 as mentioned in the inset.}
 \end{figure}

For comparison with experimentally determined dipole moments one has
to take into account the dependence on the distance between both
minima of the DWP. As already discussed above the distance for TLS
is much smaller than that for the average DWP found in our
simulations. This is a direct consequence of the generic
correlations between asymmetry and distance. Using the
parametrization method, sketched above, it is possible to calculate
the distribution of TLS from the distribution of DWP. For the
present analysis we are more specific with the definition of TLS.
Beyond a near-degeneracy of the lowest two eigenstates (energy
difference $ < 2$ K) of the TLS we also require that the relaxation
time $\tau$ is smaller than 5s.$^{17}$ This means that the TLS is
active on the time scale of typical experiments. This additional
criterion in particular excludes DWP with very high barriers and/or
large distances. This reduces the values of $d$ for the relevant TLS
even further. The results for \dwpnd\ are shown in
Fig.\ref{SiO2_d_distr_various}.

Using the parametrization method we first have recalculated the
distribution of DWP, compatible with our limiting values
$d_{max},\Delta_{max}$. This is the right curve in
Fig.\ref{SiO2_d_distr_various}. It is, of course, very similar to
the original density distribution in Fig.\ref{all_d_distr}. Using
the additional restriction $E < 2$ K one obtains a set of DWP which
is shifted towards smaller values of $d$. The final set of DWP with
the additional restriction $\tau < 5$s basically excludes DWP with
$d > 2 $\ang. The average distance of these TLS is $d_{TLS} =
1.2$\ang. Thus together with Fig.\ref{dipole_vs_d} one may estimate
a dipole moment of 0.65 D (see Tab.\ref{dipole}) for TLS in the
non-defect case which is close to the value of 0.6 D, reported in
Ref.$^{26}$.

 \begin{table}
    \begin{tabular}{|l||c|c|c|} \hline
               &$d_{TLS}$&  $\zeta$ & M  \\
          \hline
      \tlsnd\  &   1.20 \ang & 0.1 & 0.65 D \\
      \tlsod\    &  0.93 \ang  & 0.3 & 3.5 D \\
      \tlsli\   & 0.83 \ang & 1.0 & 5.0 D\\\hline
    \end{tabular}
  %\end{minipage}
  \vspace*{0.5mm}
  \caption{
 \label{dipole}The dipole moments for the different types of TLS. Included
 are the reduction factors $\zeta$  due to correlated motion of the participating
 atoms.}
  \end{table}

We have performed the same analysis for the other two cases. The
results for the values of $d_{TLS}$ and the dipole moment are given
in Tab.\ref{dipole}. Interestingly, the experimental value for
OH-defects is approx. 4 D which is close to the value for \tlsli\ as
well as \tlsod. Thus, it appears that OH-defects and Li-defects have
a similar transition mechanism during their tunneling motion.

In recent work Eq.\ref{eqmallg} has been derived, using a somewhat
different notation.$^{42}$ The derivation was based on the ideas of
the random first order transition theory. The parameter $\zeta$ was
empirically  introduced as a phenomenological constant ($\zeta
\approx 0.1$) which is needed  to recover the experimentally
determined dipole moment of \tlsnd. It has been postulated that
$\zeta q$ is the effective partial charge, stemming from a Coulomb
charge on a bead, which is much smaller than the electron charge
$e$.$^{42}$ In contrast, in the present work we can show that the
smallness of  $\zeta$ can be fully explained by the complex motional
mechanism of the TLS transitions in silica.

\section{Discussion and Summary}

Via extended computer simulations we have analyzed the properties of
the DWP in silica and lithium silicate. Most importantly, the
typical limitations of computer simulations, i.e. finite-time
(resulting in fast quenching rates) and finite-size effects, do not
hamper the characterization of the DWP within statistical
uncertainties. The latter point can be concluded from the fact that
the DWP properties do not depend on the energy of the initial glassy
configuration and, in general, for small systems the energy is most
relevant for predicting the dynamic properties.$^{31}$ This is
analogous to the case of BMLJ. We should note in passing that for
the BMLJ system properties like the vibrational density depend on
the energy of the configuration and thus on the cooling rate.$^{43}$
In contrast, for silica even the vibrational properties are to a
large extent independent of energy.$^{44}$ A quantitative estimation
of the number of DWP was possible for the BMLJ system and the
\dwpnd. In contrast, for \dwpod\  and \dwpli\ a complete search for
DWP was not possible within accessible computer times. However, one
could obtain reasonable lower bounds for the number of DWP.

The lithium ions in lithium silicate may be regarded as a prototype
system for extrinsic defects, immersed in a pure glass former. The
DWP with lithium as central particle, i.e. the \tlsli\, mainly
correspond to single lithium transitions. On a qualitative level the
situation is similar to the case of the small B-particles in the
BMLJ system which may be regarded as defects being added to the
larger A-particles. Also in this case the average DWP resembles a
single-particle transition (this holds even more for a small test
particle in the BMLJ system). Furthermore, in both cases the
probability to form DWP is significantly higher than in the
remaining glass (A-particles in BMLJ and silica in lithium
silicate). Thus extrinsic defects are very efficient in prompting
the formation of DWP.

The transitions in \dwpnd\ and \dwpod\ as well as the A-type DWP in
the BMLJ system involve the displacement of a somewhat larger number
of particles as reflected by larger values of the participation
ratio. This spatial extension, however, is still relatively small as
compared to some general predictions about the nature of TLS.$^{5}$

A closer analysis shows that \dwpnd\ and \dwpod, i.e. DWP in a
configuration without defects and with defects, respectively, behave
very differently. The transition dynamics for \dwpnd\ can be
characterized as coupled rotations of SiO$_4$-tetrahedra. In
particular this gives rise to a very small value of the dipole
moment of this transition, i.e. a very small effective charge
transport. In contrast, for \dwpod\ the dynamics is largely
determined by defect-specific modes which effectively give rise to a
much larger dipole moment. One may speculate that this is the reason
why on average the potential barrier for \dwpod\ is  much larger
than that of \dwpnd. A significant charge transport involves a major
reorganization of the whole system which typically will be connected
with a significant activation energy due to the dominant character
of the Coulomb energy.

To which degree do our simulations explain the results of
low-temperature experiments for vitreous silica?

{\it Defect-free TLS}: As already mentioned above we find approx. $4
\cdot 10^{-6}$ TLS/tetrahedron without defects. As discussed in
Ref.$^{25}$ this translates into an effective density of TLS
$\bar{P}$, accessible from acoustic experiments, which within a
factor of 3 agrees with the experimentally observed value for
Suprasil W.$^{26}$

{\it Dipole moments}: The dipole moment for \tlsnd\ is close to the
experimental value, obtained for Suprasil W.$^{26}$ Increasing the
concentration of OH, i.e. using Suprasil I, one obtains a second
contribution with a dipole moment which is larger by nearly one
order of magnitude.$^{26}$ Naturally, it has been related to the
tunneling of the OH-impurities. Interestingly, similar dipole
moments are observed for \tlsli\ and \tlsod. Since TLS with such
large dipole moments are absent in Suprasil W one may conclude that
\tlsod\ do not play a major role.

{\it Relevance of intrinsic defects}: The previous conclusion can be
directly checked by analyzing the estimated number of \tlsod. It
would be of the same size as the \tlsnd\ from the undistorted silica
network if there exists at least one intrinsic defect per 200
SiO$_4$-tetrahedra (please have in mind the uncertainty in the
determination of the number of \dwpod\ and \dwpli as discussed
above). Analysis of molecular dynamics simulations of silica, using
the metabasin approach, suggests that around $T_g$ there is roughly
1 silicon defect per 300 tetrahedra (data not shown). A different
way of extrapolation for BKS-data yields a significantly smaller
fraction of defects.$^{34}$ Furthermore, density functional
calculations in general lead to a smaller number of defects than
using the BKS-potential.$^{45}$ To the best of our knowledge no
definite experimental information is available for the number of
defects in pure silica. Thus, combining our results for the
probability of generating \tlsod\ with the (somewhat uncertain)
absolute number of intrinsic defects, the numerical results are at
least consistent with the experimentally observed absence of \tlsod\
(using the conclusion from above).

{\it Relevance of extrinsic defects}: The number of \tlsli\ would be
close to that of \tlsnd\ if there is one lithium ion per 125
tetrahedra. For Suprasil I (1200 ppm OH-defects by weight) one thus
has one OH-molecule per 240 SiO$_4$-tetrahedra. Experimentally it
has been found via dielectric echo experiments$^{26}$ as well as
specific heat experiments$^{46}$ that the number of extrinsic TLS is
half the number of intrinsic TLS. Thus an equal contribution of
extrinsic and intrinsic TLS would require one OH-molecule per
240/2=120 tetrahedra which is close to the value we obtained for an
equal contribution of \tlsli\ to \tlsnd. Of course, this perfect
agreement is to some extent accidental, given the different nature
of lithium and OH-defects and the non-completeness of the search. In
any event, the order of magnitude seems to be fully compatible.

{\it Coupling to acoustic modes:} It is estimated from experiments
that the deformation potential is approx. 40\% smaller for extrinsic
TLS than for intrinsic TLS (more specifically,  \tlsnd, according to
the discussion above). This result is obtained, first, from the
relaxation of electric echoes$^{26}$ and, second, from the
comparison of the specific heat and the thermal conductivity for
Suprasil I and Suprasil W. The difference between both probes is
much smaller for the thermal conductivity.$^{47}$ It is known from
theoretical considerations that the deformation potential scales
with the average distance $d$.$^{9}$ The different values of $d$ in
Tab.\ref{dipole} for \tlsli\ as compared to \tlsnd\ suggest that
indeed the deformation potential for extrinsic TLS may be
significantly smaller than for intrinsic TLS.

In summary, present-day computer simulations are able to reveal many
microscopic properties of two-level systems in glasses in the Kelvin
regime and enable a quantitative comparison with experimental data.
Via the complementary information from theory and experiments a
detailed knowledge about the underlying nature of tunneling systems
becomes accessible. In particular the magnetic field dependence of
polarization echo experiments may be promising to yield further
insight about the microscopic nature of TLS from the experimental
side
.$^{48,49}$ In any case, based on the additional information
from this type of simulations the tunneling systems need no longer
be considered as phenomenological entities.

\begin{acknowledgments}
We like to thank H. Lammert, A. Saksaengwijit and K. Trachenko for
fruitful discussions and the International Graduate School of
Chemistry for funding. Furthermore we would like to thank R.J.
Silbey for the initial ideas of this project.
\end{acknowledgments}

\newpage

\bibliographystyle{jpc}
%\bibliography{reference2}% Produces the bibliography via BibTeX.

\end{document}